\documentclass{aa}  

\usepackage{graphicx}
\usepackage{txfonts}
\usepackage{natbib}
\bibpunct{(}{)}{;}{a}{}{,} 

\usepackage{siunitx}
\usepackage[version=4]{mhchem} 
\usepackage[many]{tcolorbox}
\usepackage{chngcntr}

\usepackage		[hidelinks]             {hyperref}
\usepackage[capitalise] 							{cleveref}
\crefformat{figure}{Fig.\,#2#1#3}
\Crefformat{figure}{Figure\,#2#1#3}

\newcommand{\kb}{k_\mathrm{B}}
\newcommand{\Gcons}{\mathrm{G}}

\newcommand{\RB}{R_\mathrm{B}}

\newcommand{\Rp}{R_\mathrm{pl}}
\newcommand{\Rearth}{R_\oplus}
\newcommand{\RH}{R_\mathrm{H}}

\newcommand{\rhod}{\rho_\mathrm{d}}
\newcommand{\rhop}{\rho_\mathrm{p}}

\newcommand{\Mp}{M_\mathrm{pl}}
\newcommand{\Mearth}{M_\oplus}
\newcommand{\Msun}{M_\odot}
\newcommand{\Mstar}{M_*}

\newcommand{\Mpdot}{\dot{M}_\mathrm{pl}}

\newcommand{\totdev}[2]{\frac{\mathrm{d} #1}{\mathrm{d} #2}}

\newcommand{\pardev}[2]{\frac{\partial #1}{\partial #2}}
\newcommand{\devd}{\mathrm{d}}

\DeclareSIUnit{\yr}{yr}
\DeclareSIUnit\au{au}
\DeclareSIUnit\micron{\micro m}
\DeclareSIUnit\cm{cm}
\DeclareSIUnit\ppm{ppm}
\DeclareSIUnit\grav{G}
\DeclareSIUnit\bar{bar}

\begin{document}
\title{Vapor equilibrium models of accreting rocky planets demonstrate direct core growth by pebble accretion}
\author{Marie-Luise Steinmeyer 
          \inst{1}\thanks{steinmeyer\_ml@yahoo.com}
          \and
          Anders Johansen \inst{1,2}
          }
\institute{Center for Star and Planet Formation, Globe Institute, University of Copenhagen, {\O}ster Volgade 5-7, 1350 Copenhagen, Denmark
        \and 
             Lund Observatory, Department of Astronomy and Theoretical Physics, Lund University, Box 43, 221 00 Lund, Sweden
             }

   \date{Received 21/12/2023; accepted 11/02/2024}
\titlerunning{Vapor equilibrium models of accreting rocky planets}
\authorrunning{Steinmeyer \& Johansen 2024}
\abstract
   {The gaseous envelope of an accreting rocky planet becomes hot enough to sublimate silicates and other refractory minerals. 
    For this work, we studied the effect of the resulting envelope enrichment with a heavy vapor species on the composition and temperature of the envelope. For simplification, we used the gas-phase molecule SiO to represent the sublimation of silicate material. 
    We solved the equilibrium structure equations in 1D for planets in the mass range of $0.1$ to $3\,\Mearth$. The convective stability criterion was extended to take the stabilizing effect of the condensation of SiO clouds into account. We assumed that the envelope is both in hydrostatic equilibrium and in vapor equilibrium with the underlying magma ocean. This means that pebbles do not undergo sublimation in the envelope and therefore survive until they plunge into the magma ocean.
    We find that the emergence of an inner radiative region, where SiO condensation suppresses convection, increases the pressure and temperature in the inner envelope compared to pure \ce{H2}/\ce{He} envelopes once $\Mp \gtrsim 0.3\,\Mearth$. For $\Mp>0.75\,\Mearth$, the temperature and pressure close to the surface reach the supercritical point of SiO. The amount of SiO stored in the envelope is lower than the total planet mass for low mass planets. However, for $\Mp>2.0\,\Mearth$, all accreted pebble material must contribute to maintain the vapor equilibrium in the envelope. Therefore, the non-vapor mass of the planet ceases to increase beyond this threshold. Overall, our vapor equilibrium model of the planetary envelope allows for direct core growth by pebble accretion up to much higher masses than previously thought.
  
  }

   \maketitle
%
\section{Introduction}
Around one half of all Sun-like stars are thought to be orbited by planets with radii between 0.5 and 1.5 $R_\oplus$  \citep{2013Dressing,2015Silburt,2021Bryson}. The mass-radius relationship for this type of planet is consistent with a rocky composition similar to Earth \citep{2015Rogers,2015Wolfgang,2019Zeng}. Rocky planets are traditionally thought to form by giant impacts between planetary embryos after the dissipation of the protoplanetary disks \citep{2009Raymond,2020Raymond,2023BatyginMorbidelli}. However, rocky planets may also form already in the protoplanetary disk phase by pebble accretion \citep{2015Levison,2015Johansen,2019Lambrechts,2021Johansen}. In this scenario, the planets grow during the lifetime of the protoplanetary disk by accreting small solids, referred to as pebbles. These pebbles feel the drag of the surrounding gas, which increases their accretion cross section compared to the pure gravitational cross section \citep{2010Johansen,OrmelKlahr2010,Lambrechts2012}. Once a protoplanet reaches lunar mass, it starts to acquire a hydrostatic spherical envelope \citep{2012Ikoma,2014Lee}. This envelope is characterized by an increase in density compared to the local gas density in the disk. The composition of the envelope is dominated by \ce{H2} and \ce{He} \citep{2012Ikoma}.

The release of gravitational potential during the pebble accretion process heats the envelope from the bottom. Due to their small size, the pebbles are in thermal equilibrium with the envelope. Recent works have shown that the temperature in the envelope is high enough for the sublimation of pebbles to play an important role in the growth process \citep{2017Alibert,2018Brouwers,2023Steinmeyer}. Already \citet{2017Alibert} proposed that the sublimation of pebbles in the envelope leads to an end of direct growth of planets by pebble accretion as the released vapor is recycled back into the disk. However, hydrodynamical simulations have found silicate vapor recycling to be inefficient \citep{2023Wang}. Furthermore, \citet{2017Alibert} did not take the change of composition of the envelope due to pebble sublimation into account. 

In contrast, \citet{2021Ormel}, building on  \citet{2018Brouwers} and \citet{brouwers2020planets}, modeled the evolution of a planet including the enrichment of the envelope and the time-dependent heat transport. In their model, an outer region saturated in \ce{SiO2} resides on top of a possibly undersaturated region. However, this model does not take into account that rapidly formed rocky planets undergo a magma ocean phase during their formation by pebble accretion \citep{2022Olson,2023Johansen}. Furthermore, chemical equilibrium models show that the dominant gas species of \ce{Si} in equilibrium with molten silicates (\ce{MgSiO3} or \ce{Mg2SiO4}) is \ce{SiO} and not \ce{SiO2} \citep{2012Schaefer, 2020Herbort}. Similarly, \citet{2007Melosh} found that the dominant gas species after vaporization of \ce{SiO2} is \ce{SiO}. We ignore the contribution of \ce{O2} and \ce{O} in this work.

In this paper, we therefore analyze a vapor equilibrium model to represent the envelopes of accreting rocky planets close to the ice line in the mass range of $0.1\,\Mearth$ to $3\,\Mearth$. At this location, water ice sublimates and is recycled back into the protoplanetary disk even at low planet masses \citep{2021Johansen}. Therefore, this paper focuses on the accretion and sublimation of silicate pebbles. 
We want to point out three key differences to \citet{2021Ormel}. Firstly, we focus on equilibrium states and do not evolve our model in time. Secondly, we represent the sublimated silicates by \ce{SiO} instead of \ce{SiO2}. Thirdly, the underlying magma ocean acts as the major source of \ce{SiO} in our model. Therefore, the envelope is assumed to always be saturated in \ce{SiO}. This extension is based on recent models of the envelopes of super-Earths and sub-Neptunes \citep{2022Markham,2022MisenerSchlichting}.

Since SiO is heavier than the surrounding \ce{H2}/He gas, condensation of SiO has a stabilizing effect on the envelope, leading to the buildup of an inner  radiative region where the SiO contents rise to provide a significant fraction of the pressure \citep{1995Guillot,2017Leconte,2022MisenerSchlichting}. We find that this inner radiative region first appears for $\Mp \approx 0.3\,\Mearth$. The partial pressure of \ce{SiO} increases steeply with temperature. Below the inner radiative region, the increasing temperature thus leads to complete dominance of SiO in supporting the hydrostatic equilibrium. In the innermost region, the stabilizing effect of SiO condensation is gone, and the envelope becomes convective again. The main product of this paper is our calculation of the mass needed to keep the envelope saturated in SiO. We find that it is lower than the difference between two consecutive planet masses for planets with $\Mp<2.0\,\Mearth$. This implies that direct core growth by pebble accretion is possible up to masses well above one Earth mass.

The paper is organized as follows: We first describe the model for the envelope structure and the treatment of SiO vapor in \cref{sec:model}. In \cref{sec:envstruc} we discuss the effect of the envelope enrichment on the envelope structure. We also test if the treatment of the opacity and the pebble accretion rate influence the envelope structure significantly. We discuss the implication of SiO enrichment on the growth of the rocky planets in \cref{sec:planetarygrowth}. \Cref{sec:limitations} discusses the main limitations of the model. We conclude with a summary of the paper in \cref{sec:summary}.
\section{Envelope model}
\label{sec:model}
\subsection{Structure equations}
We calculate the envelope structure in 1D of a planet with mass $\Mp$ under the assumption that the envelope is in hydrostatic balance and spherically symmetric. The radial structure of the envelope thus follows the standard structure equations of hydrostatic balance and thermal gradient \citep[e.g][]{2013Kippenhahn},
\begin{subequations}
\begin{align}
    \totdev{P}{r}&= - \frac{Gm}{r^2} \rho\\
    \totdev{T}{r}&= \nabla \frac{T}{P} \totdev{P}{r}\label{eq:tempgrad},
\end{align}
\label{eq:structure}
\end{subequations}
\noindent where $\rho$, $P$, and $T$ are the density, gas pressure, and temperature respectively, and $\nabla \equiv \partial \ln T/\partial \ln P$ is the logarithmic temperature gradient. The mass enclosed at distance $r$ from the surface of the planet is given by $m$ and $G$ is the gravitational constant. 

In the case where envelope pollution is not taken into account, the gradient in \cref{eq:tempgrad} is given by the minimum of the radiative and convective gradient (Schwarzschild criterion). The radiative temperature gradient is given by
\begin{equation}
    \nabla_\mathrm{rad} \equiv \frac{3\kappa P}{64\pi G \Mp \sigma T^4}L,
    \label{eq:nabrad}
\end{equation}
where $\kappa$ is the opacity in the envelope and $\sigma$ the Stefan-Boltzmann constant. The luminosity of the planet comes from the accretion process through the expression
\begin{equation}
    L = \frac{\Gcons M \Mpdot}{R_\mathrm{rel}}.
    \label{eq:Lacc}
\end{equation}
Here, $\Mpdot$ is the pebble accretion rate. The accretion energy is released at the surface of the planet and $R_\mathrm{rel}=\Rp = (3\Mp/(4\pi \rhop))^{1/3}$. We assume that the density of the planet corresponds to the density of the uncompressed Earth, $\rhop=4050\,\si{\kilogram \per \m \cubed}$ \citep{2006Hughes}. 

The opacity plays an important role in setting the envelope structure and depends on the temperature, pressure, and composition of the envelope. The main opacity sources in the envelope are the gas molecules, pebbles, and dust grains. However, the contribution of each source remains poorly constrained. We describe the molecular gas opacity $\kappa_g$ by the analytic fit from \citet{2014Freedman} with a solar gas metallicity and the opacity coming from the dust with the power law from \citet{1994Bell}. The latter is given by
\begin{equation}
    \kappa_p = 0.1 \times T^{1/2} \,\si{\cm \squared\per\gram}.
\end{equation}
The total opacity in the nominal case is then
\begin{equation}
    \kappa = \kappa_g +  \kappa_p. \label{eq:opacity}
\end{equation}
We discuss the effect the treatment of the opacity has on the structure of the envelope in \cref{ssec:opacity}.
\subsection{Treatment of silicate vapor}
\label{ssec:silicatevapor}
\begin{figure*}[h]
    \sidecaption
    \includegraphics[width=0.7\textwidth]{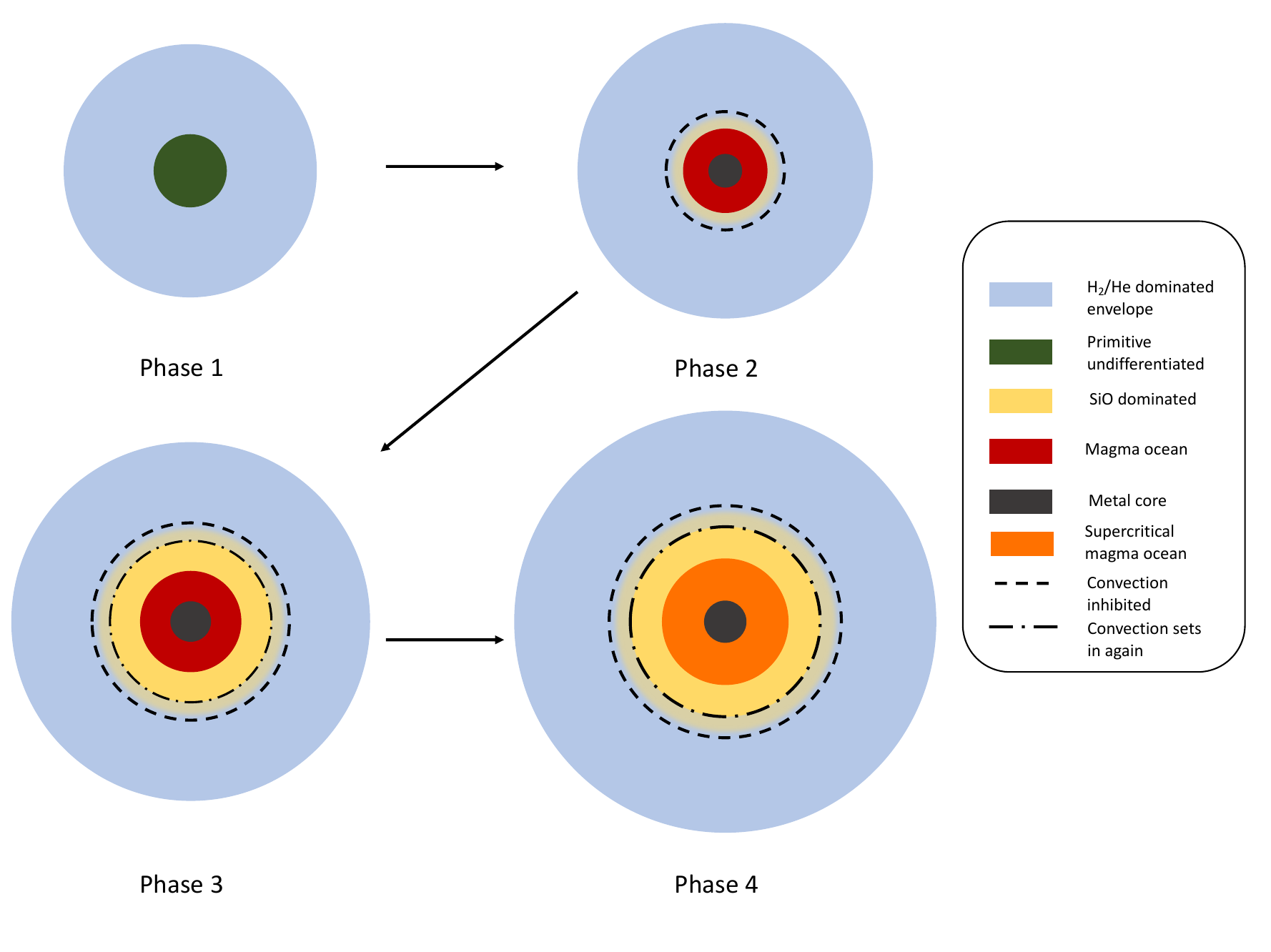}
    \caption{Illustration of the different phases of the planet and envelope during the growth process. In the beginning, the undifferentiated planet is surrounded by a \ce{H2}/He rich envelope (Phase 1). In Phase 2 the planet has differentiated into a metallic core and a magma ocean. Enough silicates sublimate in the atmosphere to inhibit convection in this phase. In Phase 3, the region close to the surface of the planet is dominated by SiO and becomes convective again. The planet becomes hot enough to form a supercritical magma ocean in Phase 4.} 
    \label{fig:envstrucmod}
\end{figure*}
Initially, the envelope has the same composition as the surrounding protoplanetary disk and consists of a mixture of hydrogen and helium (\ce{H2}/He). As the planet grows, it will melt and differentiate into an iron-rich core and a magma ocean \citep{2022Olson,2023Johansen}. The envelope is assumed to be in vapor equilibrium with this magma ocean at all times. Therefore, silicate vapor evaporated from the magma ocean will become stable in the envelope. For simplicity, we assume that the representative gas-phase molecule in equilibrium with the magma ocean is \ce{SiO}. The partial pressure of SiO then follows the saturated vapor pressure of SiO. We take the expression from \citet{2012FegleySchaefer} and \citet{2013VisscherFegley},
\begin{equation}
    P_{\ce{SiO}}=P_\mathrm{svp} = A 10^{-B/T},
    \label{eq:PSiO}
\end{equation}
where $P_\mathrm{svp}$ is given in bar, $A= 10^{8.203}\,\si{\bar}$ and $B=25898.9\,\si{\K}$.

The adiabatic temperature gradient of an atmosphere with a condensable species, in this case \ce{SiO}, is described by the moist adiabatic gradient \citep{2017Leconte,2022MisenerSchlichting},
\begin{equation}
    \nabla_\mathrm{ad} = \frac{\kb}{\mu} \frac{1 + \frac{P_{\ce{SiO}}}{P_\mathrm{bg}}\pardev{\ln P_{\ce{SiO}}}{\ln T}}{c_p +  \frac{P_{\ce{SiO}}}{P_\mathrm{bg}}\frac{\kb}{\mu}\left(\pardev{\ln P_{\ce{SiO}}}{\ln T}\right)^2}.
    \label{eq:moistad}
\end{equation}
Here $\kb$ is the Boltzmann constant and $P_\mathrm{bg} = P - P_{\ce{SiO}}$ is the partial pressure of \ce{H2} and He. The heat capacity of a gas with mean molecular weight $\mu$ and an adiabatic index of $\gamma$ is
\begin{equation}
    c_p = \frac{\kb}{\mu} \frac{\gamma}{\gamma-1}.
    \label{eq:cp}
\end{equation}
The dominant gas species are the diatomic molecules \ce{SiO} and \ce{H2}, which both display $\gamma=1.4$ under the assumption of ideal gas behavior. If $P_{\ce{SiO}}\ll P_\mathrm{bg}$, \cref{eq:moistad} follows the dry adibat 
\begin{equation}
    \nabla_\mathrm{ad} = \frac{\gamma-1}{\gamma}.
\end{equation}
The local mean molecular weight of the enriched envelope is given by
\begin{equation}
    \mu = \frac{\mu_\mathrm{bg}P_\mathrm{bg} + \mu_{\ce{SiO}}P_{\ce{SiO}}}{P}, 
    \label{eq:meanmolecweight}
\end{equation}
with $\mu_\mathrm{bg}=2.34$ and $\mu_{\ce{SiO}}=44$ given in atomic mass units. 

According to \cref{eq:PSiO} the hotter the envelope is, the more \ce{SiO} will be contained in the envelope. This increase in $\mu$ with temperature has a stabilizing effect on the envelope \citep{1995Guillot,2017Leconte}. If the mass mixing ratio
\begin{equation}
    q=\frac{\mu_{\ce{SiO}}P_{\ce{SiO}}}{\mu_\mathrm{bg} P_\mathrm{bg}}
\end{equation} 
reaches a threshold value $q_\mathrm{th}$, convection is inhibited. The threshold mixing ratio is given by \citet{1995Guillot} and \citet{2017Leconte} as
\begin{equation}
    q_\mathrm{th} = \frac{1}{\left(1-\frac{\mu_{\ce{H}}}{\mu_{\ce{SiO}}}\right)\pardev{\ln P_{\ce{SiO}}}{\ln T}}.
    \label{eq:qcrit}
\end{equation}
This criterion is different from the \citet{1947Ledoux} criterion for stability in a medium with a mean molecular weight gradient. In the case of the Ledoux criterion, the composition of a moving gas element is fixed while the local background mean molecular weight changes, whereas \cref{eq:qcrit} takes the change of mean molecular weight of a moving gas element due to condensation into account. In regions where $q<q_\mathrm{th}$, the temperature gradient follows \cref{eq:nabrad}. However, if the local mass mixing ratio reaches a value $q=q_\mathrm{max}\gg 1$, the local envelope mainly consists of SiO and the stabilizing effect no longer holds. Thus, for $q>q_\mathrm{max}$ convection sets in again. In this region the saturated vapor pressure of SiO dominates, $P_{\ce{SiO}}\gg P_\mathrm{bg}$, and the moist adiabatic gradient follows the coexistence curve of \ce{SiO}
\begin{equation}
    \nabla_\mathrm{ad} = \pardev{\ln T}{\ln P_{\ce{SiO}}}.
\end{equation}

The temperature and pressure close to the surface will eventually become hot enough to reach the critical point of \ce{SiO} at temperature $T_\mathrm{crit} \approx 6600\,\si{K}$ and pressure $P_c \approx 1400\,\si{\bar}$ \citep{2018XiaoStixrude}. We therefore treat the region with $T>T_c$ as the outer mantle or supercritical magma ocean \citep{2018Bodenheimer,2022Markham}. The radius of the supercritical magma ocean, $R_\mathrm{crit}$, is defined as the radius where $T>T_\mathrm{crit}$ for the first time. In this case, the accretion heating is no longer released at the surface of the planet but on top of the supercritical magma ocean, which translates to $R_\mathrm{rel}=R_\mathrm{crit}$ in \cref{eq:Lacc}. We treat the region below $R_\mathrm{crit}$ as part of the mantle of the planet and not the envelope, similar to \citet{2018Bodenheimer}. A consequence of this is that the temperature at the bottom of the envelope never exceeds $T_c$. 
\begin{figure*}[h!]
    \centering
    \includegraphics[width=\textwidth]{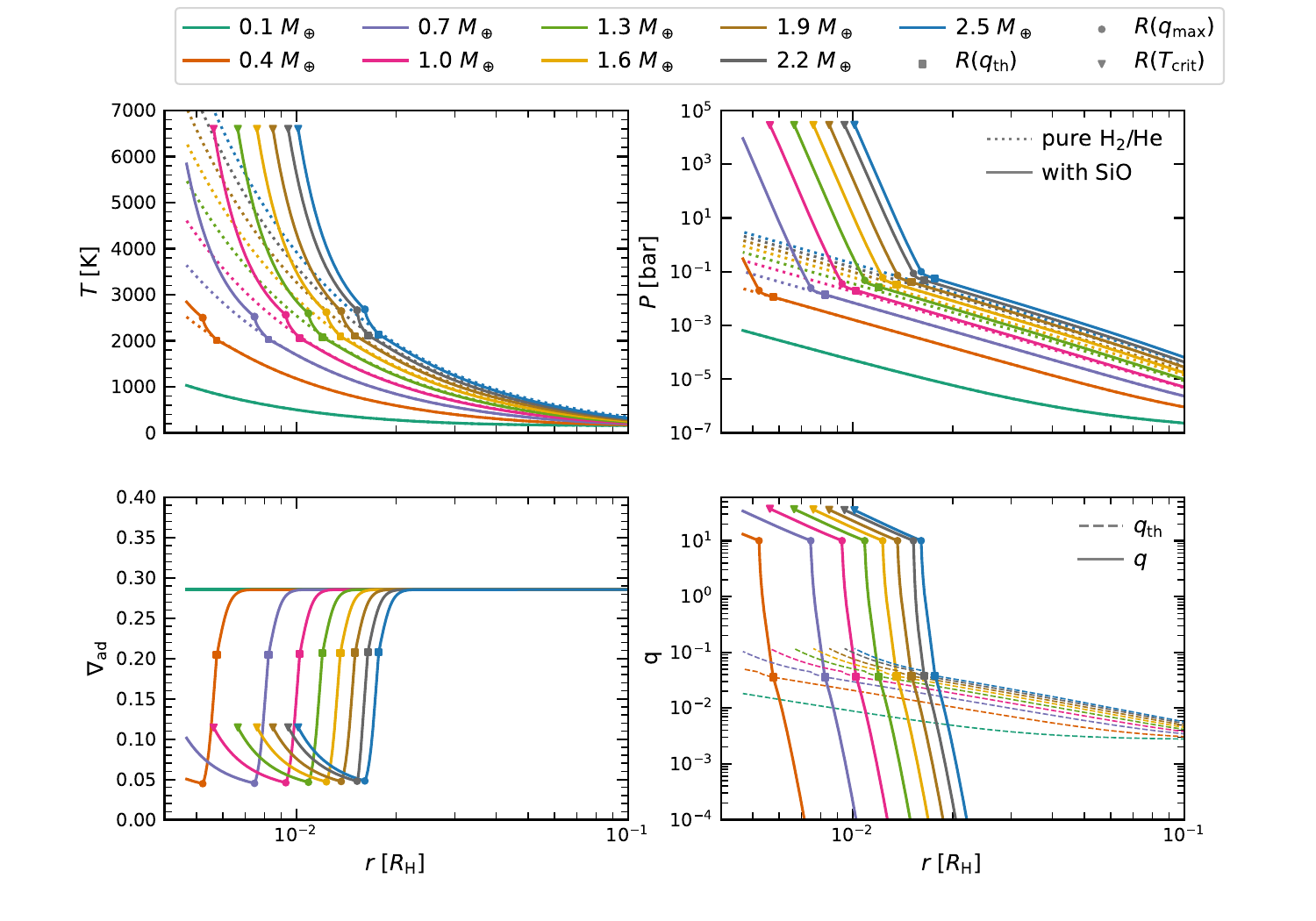}
    \caption{Envelope profiles of a selected set of planet masses accreting at $\dot{M}_{\mathrm{p}}=10^{-6}\,\Mearth\si{\per\yr}$. The distance to the core is given in units of the Hill radius. The color corresponds to the mass of the planet. The top row shows the temperature (left) and pressure (right). Solid lines include the enrichment with SiO while dashed lines correspond to a pure \ce{H2}/He envelope. The bottom row shows the moist adiabatic gradient including the latent heat from SiO condensation (left) and the mass mixing radio (right). The critical mass ratio is shown by dashed lines. In all four plots,  the squares indicate the start of the inner radiative region and the filled circles the end. Planets larger than $0.75\,\Mearth$ possess a supercritical magma ocean. The surface of this ocean is marked by the triangles.}
    \label{fig:envelevMdot1e6}
\end{figure*}
\begin{figure}[h!]
    \centering
    \includegraphics[width=\hsize]{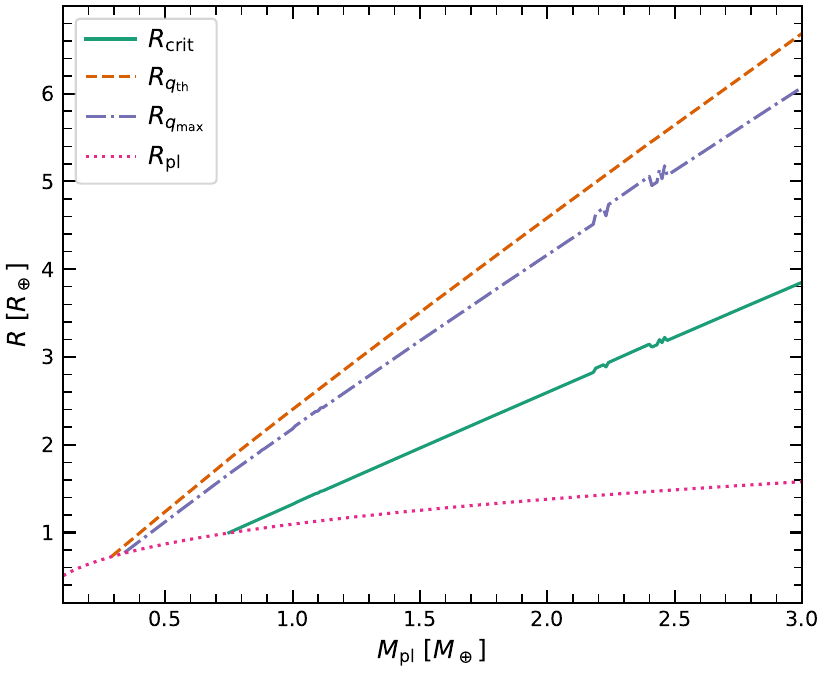}
    \caption{Key radii as a function of the planet mass. The dashed orange line indicates the location where $q>q_\mathrm{th}$ for the first time, while the purple dash-dotted line shows the location where $q>q_\mathrm{max}$ for the first time. The solid green line indicates the surface of the supercritical magma ocean, and the dotted pink line the radius of the planet without the supercritical magma ocean. For planets with $\Mp \geq 0.75\,\Mearth$, the accretion heating from the incoming pebbles is released at the surface of this supercritical magma ocean. The feature between $2.1\,\Mearth$ and $2.5\,\Mearth$ in $R_\mathrm{crit}$ and $R_{q_\mathrm{max}}$ is due to slight discontinuities in the solution, caused by the finite numerical grid.}
    \label{fig:RTcrit}
\end{figure}

The different phases of the planet and envelope in our model are shown in \cref{fig:envstrucmod}. In phase 1, the undifferentiated planet is surrounded by a \ce{H2}/He rich envelope. The temperature gradient in the envelope is set by $\nabla = \min(\nabla_\mathrm{ad},\nabla_\mathrm{rad})$. The temperature in the inner envelope increases and the planet differentiates into a metal core and a magma ocean mantle. \ce{SiO} will start to evaporate from the magma ocean in order to keep the equilibrium between the magma ocean and the surrounding envelope. Once $q$ reaches the critical value, an inner radiative region appears (Phase 2). As the magma ocean temperature increases, more SiO enters the envelope and a SiO-dominated convective region forms beneath the inner radiative zone (Phase 3). In Phase 4, the temperature close to the surface of the planet has become hot enough to transition the magma ocean into a supercritical magma ocean state. 
\subsection{Numerical method}
We solve \cref{eq:structure} together with \cref{eq:meanmolecweight} by integrating from the outer boundary down to the surface of the planet, using a 4th order Runge-Kutta method. We set the outer boundary to the Hill radius of the planet,
\begin{equation}
    \RH = a \left(\frac{\Mp}{3M_\star}\right)^{1/3},
\end{equation}
where $a$ is the distance of the planet from the star. We take the stellar mass $\Mstar$ to be equal to $1\,\Msun$. The density and temperature at the Hill radius are set to match the surrounding protoplanetary disk, $\rho = \rhod$ and $T=T_\mathrm{d}$. We determine the step size $\devd r$ via
\begin{equation}
    \devd r_{i+1} = \min\left(c\frac{T_i}{|\devd T_i /\devd r_i|},\devd r_\mathrm{log}\right),
\end{equation}
with $c\leq 10^{-3}$ and $\devd r_\mathrm{log}$ from a logarithmically spaced grid. We chose this set up to account for the steep temperature gradients that appear in the enriched inner envelope. 
\begin{figure*}[h!]
    \centering
    \includegraphics[width=\hsize]{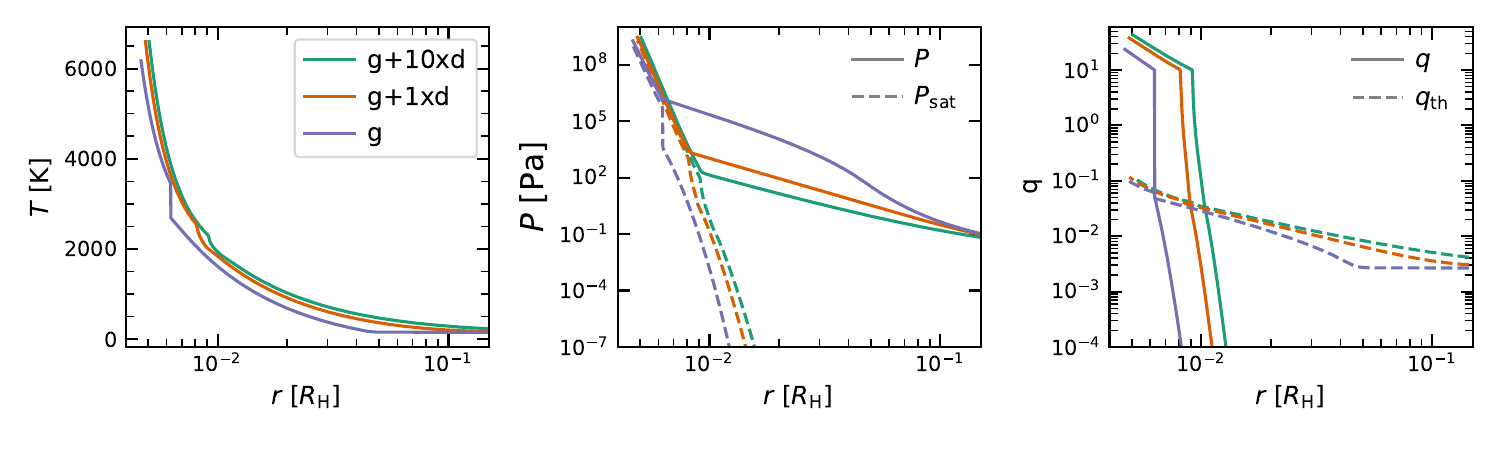}
    \caption{Comparison of the temperature, pressure, and SiO mass mixing ratio profiles of a planet with $\Mp=0.8\,\Mearth$ for three different treatments of the opacity. The model ''g+1xd'' corresponds to the nominal case. In model ''g+10xd'', the dust contribution is increased by a factor 10. In model ''g'', the gas is the only opacity source considered. Due to the larger outer isothermal region in model ''g'', the temperature in the envelope is lower than in the other models while the pressure is higher.}
    \label{fig:envstrucM08divopMpeb1e6}
\end{figure*}

Initially, we set the mean molecular weight in the envelope to be that of a mixture of $\ce{H2}$ and \ce{He}, $\mu_0=2.34\,\mathrm{m}_\mathrm{u}$. After solving for the envelope structure, we calculate the new mean molecular weight profile of the envelope via \cref{eq:moistad} and \cref{eq:meanmolecweight}. The mean molecular weight gradient determines the adiabatic gradient and consequently the temperature and pressure profile in the envelope. We then solve \cref{eq:structure} again using the new mean molecular weight profile. For each iteration, we calculate the total mass in \ce{SiO} as a function of the resulting temperature and pressure profile according to
\begin{equation}
    M_{\ce{SiO}} = \int_{R_\mathrm{rel}}^{\RB} 4\pi r^2 \rho_{\ce{SiO}} \devd r = \int_{\Rp}^{\RB} 4\pi r^2 \frac{P_{\ce{SiO}} \mu_{\ce{SiO}} m_u}{\kb T} \devd r,
    \label{eq:atmmass}
\end{equation}
as well as the new mean molecular weight profile. The outer radius in \cref{eq:atmmass} is set to $\RB$ since material outside this radius is likely replenished by \ce{H2}-rich gas from the disk \citep{2017LambrechtsLega,2018KurokawaTanigawa}. Additionally, the temperatures in the outer envelope are too low to contain a significant amount of \ce{SiO}. We repeat this process until the SiO mass of the envelope converges. On average, it took 10 iterations for each planet mass to converge.  
\section{Effect of SiO vapor on the envelope structure}
\label{sec:envstruc}
\subsection{Nominal case}
We calculate the envelope profile of planets between $0.1\,\Mearth$ and $3\,\Mearth$ with a mass step of $\Delta \Mp = 0.01\,\Mearth$. The location of the planet is kept constant at $1\,\si{\au}$ with $T_{d}=150\,\si{\K}$ and $\rho_d=1.7\times10^{-8}\,\si{\kg\per\m\cubed}$. This corresponds to a gas surface density of $\Sigma_\mathrm{g}\approx 310\,\si{\kg \per \m \squared}$, which is two orders of magnitude lower than the gas surface density at $1\,\si{\au}$ given by the Minimum Mass Solar Nebulae, $\Sigma_\mathrm{g}(\mathrm{MMSN}) \approx 1.7 \times 10^{4}\,\si{\kg \per \m \squared}$, \citep{1981Hayashi}. However, the pressure and temperature in the inner region of the envelope are relatively independent of the outer boundary condition. 
The nominal pebble accretion rate is set to $\dot{M}_{\mathrm{p}} = 10^{-6}\,\Mearth\si{\per \yr}$. \Cref{fig:envelevMdot1e6} compares the enriched envelope profiles of selected masses to the structure of pure \ce{H2}/He envelopes. The outer envelope is too cold for silicate sublimation to play an important role. As a consequence, the envelope profiles in the outer envelope of all planet masses are equal to the pure \ce{H2}/He envelopes. 

Phase 1 in which the whole envelope contains mainly \ce{H2}/He ends after approximately $\Mp \geq 0.29\,\Mearth$. At this point the surface temperature is $2000\,\si{\K}$ and the planet has therefore melted and differentiated into a metal core and a molten silicate magma ocean. Since the saturated vapor pressure is a strong function of the temperature, the envelope around low mass planets is easily saturated in SiO from the magma ocean at this stage. Therefore, the incoming pebbles continue to move through the envelope without sublimation. The temperature and saturated vapor pressure in the inner region of the envelope are high enough so that the mass mixing ratio reaches the critical value to suppress convection by SiO cloud condensation. The envelope thus develops an inner radiative region. The radiative temperature gradient in this inner region is much steeper than the adiabatic gradient, which leads to an increase in temperature compared to the \ce{H2}/He envelopes; this is clear from the top left plot in \cref{fig:envelevMdot1e6}. Increasing the planetary mass further, the envelope temperature increases and the SiO mixing ratio increases accordingly, as is seen in the bottom right plot in \cref{fig:envelevMdot1e6}. This is denoted Phase 2 in \cref{fig:envstrucmod}. Once $\Mp\geq 0.34\,\Mearth$, the mass mixing ratio becomes larger than $q_\mathrm{max}$ and an inner convective region develops (Phase 3). The total pressure in this region is completely dominated by the partial pressure of SiO, which leads to a significant increase in pressure compared to the pure \ce{H2}/He envelopes. The critical point of SiO is first reached for a planet with $\Mp=0.75\,\Mearth$. The temperature at the bottom of the envelope is constant, because the region below $R_\mathrm{crit}$ is part of mantle of the planet.
\begin{figure}[h]
    \centering
    \includegraphics[width=\hsize]{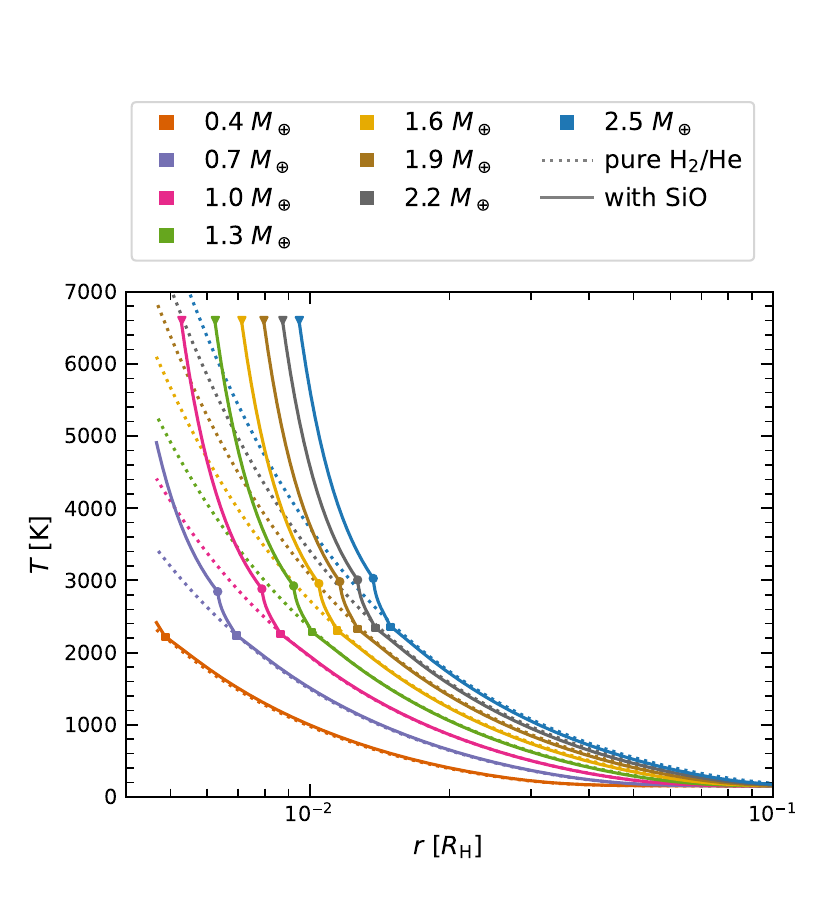}
    \caption{Temperature profiles of a selected set of planet masses and $\dot{M}_{\mathrm{peb}}=10^{-7}\,\Mearth\si{\per\yr}$. Colors and line styles have the same meaning as in \cref{fig:envelevMdot1e6}. At a given mass, the envelopes are slightly colder than in the nominal case $\dot{M}_{\mathrm{peb}}=10^{-6}\,\Mearth\si{\per \yr}$. As a result, the enriched envelopes start to differ from the pure \ce{H2}/He envelopes only after $\Mp>0.4\,\Mearth$.}
    \label{fig:envelevMdot1e7}
\end{figure}

\Cref{fig:RTcrit} shows the outer ($q=q_\mathrm{th}$) and inner edge ($q=q_\mathrm{max}$) of the inner radiative region as well as the surface of the supercritical magma ocean as a function of the planet mass. The width of the inner radiative region increases from $0.08\,\Rearth$ at $\Mp = 0.34\,\Mearth$ to $0.61\,\Rearth$ at $\Mp = 3.0\,\Mearth$. The extent of the SiO-dominated region below the inner radiative exceeds $1\,\Rearth$ for $\Mp>1.19\,\Mearth$.
\subsection{Influence of opacity}
\label{ssec:opacity}
\citet{1994Bell} assume a solid-to-gas ratio of $Z=0.01$ in micron-sized particles to calculate the dust opacity. However, the actual solid-to-gas ratio in a planetary envelope is likely higher \citep{2020Alidibthompson,2020JohansenNordlund}. We therefore compare three different opacity cases in \cref{fig:envstrucM08divopMpeb1e6}. The standard model ''g+1xd'' is given by \cref{eq:opacity} while in the model ''g+10xd'' we increase the dust contribution and use $\kappa = \kappa_g + 10 \times \kappa_p$. Lastly, in model ''g'' we calculate the envelope using only the gas opacities following \citet{2014Freedman}. For all three opacity models in \cref{fig:envstrucM08divopMpeb1e6}, the mass of the planet is $\Mp = 0.8\, \Mearth$ and the pebble accretion rate is $\dot{M}_{\mathrm{p}}=10^{-6}\,\Mearth\si{\per \yr}$. Increasing the contribution of the dust to the opacity leads to a small increase in temperature.  Therefore, $q_\mathrm{th}$ and $q_\mathrm{max}$ are reached further out in the envelope for model 'g+10xd'. The profile of the envelope changes significantly if we consider only the gas opacity. Due to the low gas opacity in the outer envelope, the envelope cools effectively and the outer isothermal region reaches further into the envelope. As a consequence, the temperatures in the envelope are lower than if the dust opacity is included. Due to the strong increase in pressure in the isothermal region, the overall pressure structure is higher in model 'g'. The critical mass mixing ratio is only reached at $0.006\,\RH$ compared to $0.009\,\RH$ for 'g+1xd' and $0.01\,\RH$ for 'g+10xd'. The temperature gradient in the inner radiative region in model 'g' is very steep. Therefore, the envelope almost immediately becomes dominated by SiO with $\Delta R_\mathrm{rad}=0.0004\,\Rearth$. 
\begin{figure}[h]
    \centering
    \includegraphics[width=\hsize]{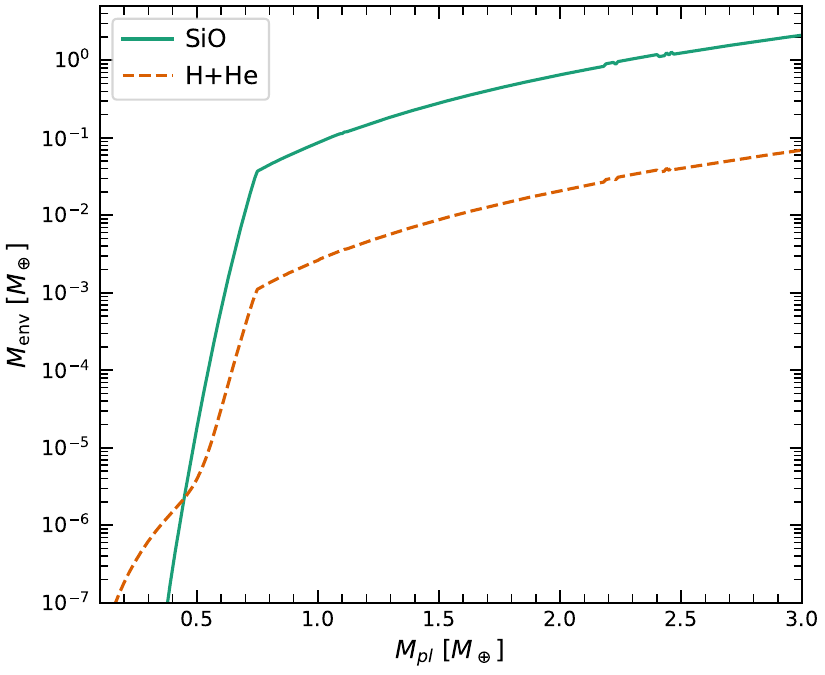}
    \caption{Total envelope mass in Earth masses as a function of the planet mass. The solid lines show the SiO mass while the dotted lines show the mass of \ce{H2}/He in the envelope. The SiO mass dominates the total mass budget for planets with $\Mp>0.6\,\Mearth$. The kink shows the start of the supercritical magma ocean after which the amount of SiO and \ce{H2}/He in the envelope increases only by heating the outer parts of the envelope.}
    \label{fig:atmmassM013}
\end{figure}
\subsection{Influence of pebble accretion rate}
For completeness, we also calculate the envelope profiles for a lower pebble accretion rate of $\dot{M}_{\mathrm{p}} = 10^{-7}\,\Mearth\si{\per \yr}$. As is shown in \cref{fig:envelevMdot1e7}, the temperatures in the envelope of a given planet mass are lower compared to the case where  $\dot{M}_{\mathrm{p}} = 10^{-6}\,\Mearth\si{\per \yr}$. This is because the accretion heating scales with the pebble accretion rate, see \cref{eq:Lacc}. Otherwise, there is no significant difference between a high and a low pebble accretion rate. 
\begin{figure}[h]
    \centering
    \includegraphics[width=\hsize]{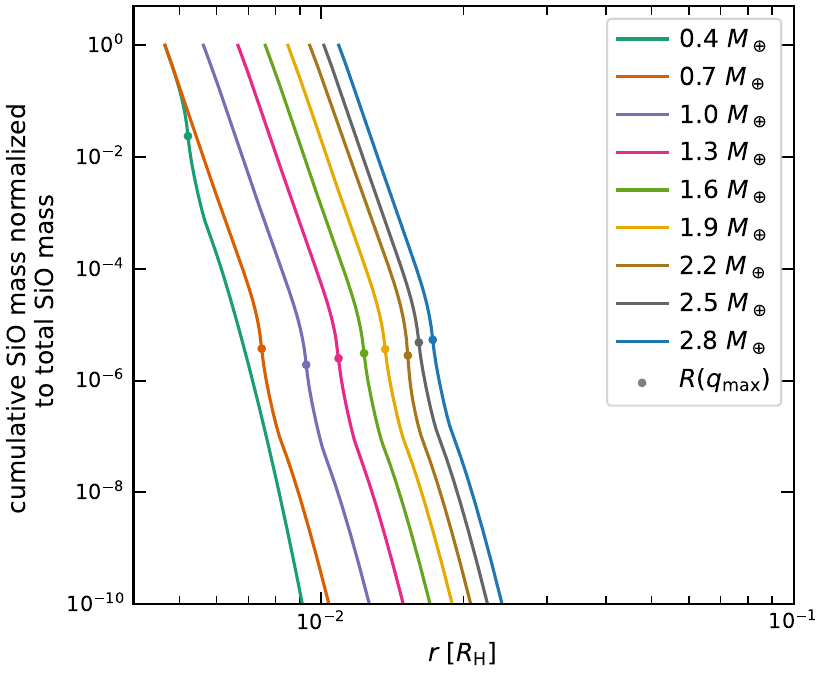}
    \caption{Cumulative mass distribution of SiO for different planet masses. The distance to the surface of the planet is given in units of the Hill radius of the respective planet. The filled circles mark the end of the inner radiative zone. Most of the SiO vapor mass is concentrated inside the radiative region. }
    \label{fig:Mcumsumg1mMd6}
\end{figure}
\section{Implication for planetary growth}
\label{sec:planetarygrowth}
\subsection{Total atmosphere mass}
\begin{figure}[h]
    \centering
    \includegraphics[width=\hsize]{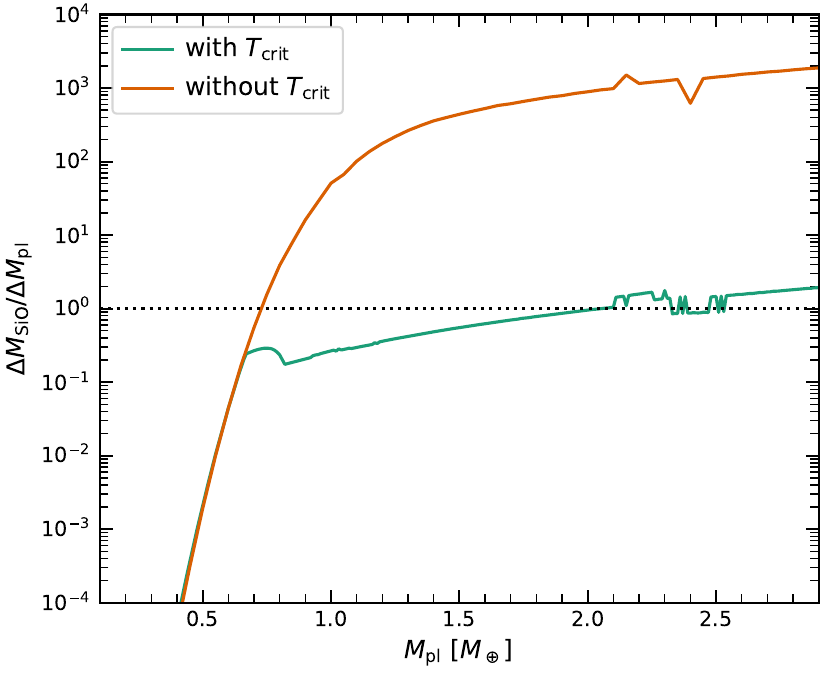}
    \caption{Ratio of the amount of SiO needed to keep the envelope saturated to the total mass added to the planet as a function of the planet mass. The green line shows the nominal model where the critical point of \ce{SiO} is considered. The orange line shows the case where the critical point is not taken into account. In the nominal case, $\Delta M_{\ce{SiO}}/\Delta \Mp$ becomes larger than unity for $\Mp\geq2.0\,\Mearth$. However, $\Delta M_{\ce{SiO}}/\Delta \Mp$ never becomes larger than 2 and the envelope is always saturated in \ce{SiO}. In the model without $T_\mathrm{crit}$, $\Delta M_{\ce{SiO}}/\Delta \Mp$ reaches unity already at $\Mp = 0.8\,\Mearth$. The key different between the two models, however, is that the amount of \ce{SiO} needed to keep the envelope saturated in the model without $T_c$ becomes larger than the total mass of the planet for $\Mp \gtrsim 1.0\,\Mearth$. In this case, the envelope is most likely undersaturated. The feature between $2.1\,\Mearth$ and $2.5\,\Mearth$ is likely caused by the inner boundary of the envelope switching between two grid cells.}
    \label{fig:dSiOMEM0114}
\end{figure}
The total mass in SiO and \ce{H2}/He as a function of planet mass is shown in \cref{fig:atmmassM013}. As expected, the total mass in the envelope is dominated by \ce{H2}/He for $\Mp<0.4\,\Mearth$. The envelope mass increases steeply until the critical point of SiO is reached at $\Mp=0.75\,\Mearth$. The temperature at the bottom of the envelope is set constant at $T_c$ for higher masses. The envelope thus only grows by heating the outer parts of the envelope. Therefore, the mass in the envelope increases weakly with planetary mass. The total mass of SiO is almost two orders of magnitude larger than the total mass in \ce{H2}/He for $\Mp>0.73\,\Mearth$.

In \citet{2023Steinmeyer}, we proposed that the sublimation of silicates and other refractory mineral species will create a layer rich in heavy gas species, such as \ce{SiO}, close to the surface of the planet. We further proposed that this layer is protected from recycling flows by an inner radiative zone. This is in line with hydrodynamic simulation of atmospheric recycling \citep{2023Wang}. We now test this hypothesis by plotting the cumulative mass distribution of the SiO in the envelope for different planet masses in \cref{fig:Mcumsumg1mMd6}. Almost all SiO is indeed concentrated below the inner radiative region. As discussed before, this is because the saturated vapor pressure is low in the outer envelope. Therefore, pebbles plunge into the magma ocean without experiencing sublimation in the envelope. The magma ocean in turn acts as a reservoir of SiO to saturate the envelope at the temperature conditions given by the planetary mass and accretion rate. 

\subsection{End of direct core growth}
It has been proposed by other authors that direct growth stops after the planet reaches a mass as low as $0.5\,\Mearth$, because all incoming pebbles are destroyed before they reach the core \citep{2017Alibert,2018Brouwers,brouwers2020planets,2021Ormel}. We therefore calculate the amount of SiO that is needed to keep the envelope saturated by calculating the change in SiO in the envelope, $\Delta M_{\ce{SiO}}$, between two consecutive masses and compare it to the accreted mass $\Delta \Mp$. The total envelope mass shows some noise for planets with masses between $2.1\,\Mearth$ to $2.5\,\Mearth$. The most likely cause for this is that the surface of the magma ocean lies at the border of two grid cells. We therefore used a boxcar average to smooth the total \ce{SiO} mass in the envelope in order to avoid negative values in $\Delta M_{\ce{SiO}}$. The results are shown in \cref{fig:dSiOMEM0114}. Similarly to the total envelope mass, the ratio $\Delta M_{\ce{SiO}}/\Delta \Mp$ increases steeply until the supercritical magma ocean forms. However, for $\Mp<0.6\,\Mearth$, $\Delta M_{\ce{SiO}}/\Delta \Mp$ is very small with a value of less than $0.01$. Once $\Mp\geq2.0\,\Mearth$, $\Delta M_{\ce{SiO}}/\Delta \Mp$ becomes larger than unity. We propose that in this case the pebbles actually sublimate in the envelope and the region where the accretion heat is released moves outward. The envelope cools to restore vapor equilibrium at the modified accretion luminosity profile. At the same time, the underlying supercritical magma ocean continues to act as a source of \ce{SiO}. This way, the envelope will maintain saturation in SiO as the accretion heat is released well above the magma ocean surface.

\citet{2021Ormel} modeled the evolution of a planet including the time evolution of the heat transport and the moist adiabat. Although this is a different approach than the vapor equilibrium model presented in this paper, we can still compare the general conclusions. \citet{2021Ormel} found that the envelope is divided in an outer region which is saturated in \ce{SiO2} and an inner under-saturated region. However, the key difference in assumptions between \citet{2021Ormel} and the model presented in this paper is the existence of a magma ocean and later a supercritical magma ocean. \Cref{fig:dSiOMEM0114} shows that if the critical point of \ce{SiO} is ignored, the amount needed to keep the envelope saturated in \ce{SiO} becomes larger than $\Delta\Mp$ already for $\Mp\geq0.8\,\Mearth$. More importantly, in the case without considering the transition to a supercritical fluid, $\Delta M_{\ce{SiO}}/\Delta \Mp>10^2$ for $\Mp \gtrsim 1.0\,\Mearth$.  In this case the envelope becomes undersaturated in \ce{SiO} as predicted by \citet{2021Ormel}. 

The main heat source of the envelope is the transformation of gravitational potential to thermal energy during the accretion process. The pebble flux could cease either due to depletion radial drift or if an outer planet opens a gap that interrupts the pebble flux \citep{2012MorbidelliNesvorny,2014Lambrechts,2024Gurrutxaga}. In any case, the envelope will begin to  cool down and the supercritical magma ocean will transition into a normal magma ocean. As the temperatures in the envelope become lower, the saturated vapor pressure of silicate decreases. As a consequence, the silicate vapor will condense and rain out on the surface of the planet \citep{brouwers2020planets,2023VazanOrmel}. 

\subsection{Long-term evolution of composition gradients}
According to structure models of Jupiter that are based on measurements by the Juno and Galileo missions, the planet contains an extended region in its deep interior that is rich in heavy elements \citep{2017WahlJuno,2019DebrasChabrierJuno,2024MilitzerHubbard}. This region is referred to as the "dilute" or "fuzzy core" of Jupiter. An outward decreasing interior composition gradient is furthermore consistent with the available measurements of Uranus and Neptune \citep{1995Marley,2000Podolak,2011HelledUranusNeptune}. These composition gradients can be seen as a remnant of compositional gradients coming from formation \citep{2020VenturiniHelled,2021Ormel}. However, all these planets have accreted a significant amount of \ce{H2}/He-gas. We find that low mass planets, $\Mp<2.5\,\Mearth$, that only accreted modest amounts of \ce{H2}/He-gas have steep compositional gradients, see \cref{fig:envelevMdot1e6}. Furthermore, the envelope is always saturated in \ce{SiO}. Hence, our results are not directly applicable to ice and gas giants. The presence of extended vapor regions in our model may nevertheless be important for understanding the dilute cores of giant planets. Adding gas accretion and mixing between vapor and accreted gas will nevertheless be needed to understand the connection between the vapor-rich cores found here and the dilute cores observed in the giant planets of the Solar System. 

\section{Limitations of the model}
\label{sec:limitations} 
\subsection{Evaporation timescale}
In this work, we assumed that the evaporation of \ce{SiO} from the magma ocean is a very efficient process. The mass loss rate due to evaporation is \citep{2019Ros}
\begin{equation}
    \dot{M}_\mathrm{evap} = 4 \pi R^2 v_\bot \rho_\mathrm{svp},
    \label{eq:evaprate}
\end{equation}
where $R$ is the radius of the planet, $v_\bot = \sqrt{k_b T/(2\pi m_{\ce{SiO}})}$ the average speed of vapor, and $\rho_\mathrm{svp} = \mu_{\ce{SiO}} P_\mathrm{svp}/(\kb T)$. Based on \cref{fig:envelevMdot1e6}, typical values for $P_\mathrm{svp}$ at the surface of the planets are on the order of $10^{3}\,\si{\bar}$, which translates to $\rho_{\mathrm{sat}} \sim 100\,\si{\kg \per \m\cubed}$. The average speed of the vapor is $v_{\bot} \sim 100 \si{\m\per\s}$. For $1\,\Mearth$, this corresponds to a mass loss rate of $\dot{M}_\mathrm{evap}\sim 10^{-7}\,\Mearth\si{\per\s}$. Therefore, the evaporation rate of the planet will be extremely fast compared to the accretion timescale even at low $\rho_\mathrm{sat}/\rho_\mathrm{p}$. For this reason, the envelope is in equilibrium with the magma ocean at any time.  

\subsection{Composition of pebbles}
The main assumption in this paper is that the pebbles are purely made out of silicates and that these silicates turn into \ce{SiO} vapor after sublimation. Pebbles that formed in an environment with solar composition have a water ice-to-rock ratio of roughly 1:1 \citep{2003Lodders}. Therefore, planets forming outside the water ice line should accrete water ice as well as silicates. However, \citet{2021Johansen} found that in the vicinity of the water ice line, water ice will start to sublimate in the envelope once the planet reaches a mass of $0.02\,\Mearth$. The resulting water vapor can easily be recycled back into the protoplanetary disk \citep{2021Johansen,2023Wang}. 

While pebbles can contain a variety of mineral species with different sublimation behavior, the fraction of both more refractory (e.g., \ce{Al2O3}) and less refractory (e.g., \ce{FeS}) species is low compared to water ice and silicates \citep{1994Pollack}. For that reason, we can assume that the incoming pebbles only contain silicates. The most likely silicates to form if the pebbles condensed out of solar composition material is enstatite (\ce{MgSiO3}) \citep{Gail1998}. Experiments have shown that the sublimation of silicates is a fast process \citep{1999Tsuchiyama,2002Tachibana}. Enstatite sublimates by forming a layer of forsterite (\ce{Mg2SiO4}) \citep{2002Tachibana}. The final sublimation products of enstatite and forsterite are atomic Mg, \ce{SiO}, and \ce{O2} \citep{2002Tachibana,2009Schaefer}. In the inner envelope the temperature and pressure are high enough for \ce{Si} and \ce{O} to be stable \citep{2007Melosh}. Since atomic Mg, Si, and O as well as \ce{O2} have a lower molecular weight than \ce{SiO}, our assumption of purely SiO gas overestimates the mean molecular weight of the heavy gas species. Furthermore, we neglected the possibility of reactions between \ce{SiO} and \ce{H2}, which can lead to the production of species such as \ce{SiH4} or \ce{H2O} \citep{2023Misener}. Nevertheless, using SiO as the key gas species is a good first approximation for the sublimation of refractory rocky material in the envelope of an accreting planet.

The molecular gas opacity scales with the metallicity of the gas \citep{2014Freedman}. The release of \ce{SiO} due to pebble sublimation in the innermost region increases the metallicity of the gas, which therefore leads to an increase of the gas opacity. However, regions where the change in metallicity is significant are anyway convective, and hence the opacity plays no role for heat transport in these regions. Therefore, we used a constant solar metallicity in order to calculate the molecular gas opacity.
\subsection{Critical point of SiO}
We set the critical point of SiO to $T_c=6600\, \si{\K}$ and $P_c=140\,\si{\mega\Pa}$ \citep{2018XiaoStixrude}. Literature values for the critical temperature are in the range of $5000$ to $15000\,\si{\K}$ \citep{2007Melosh,2012Kraus,2014Iosilevskiy,2016Connolly}. The critical point determines the onset of the supercritical magma ocean. If the critical temperature is lower than $6600\,\si{\K}$, the supercritical magma ocean will appear already at lower masses. Therefore, the amount of gaseous SiO in the envelope is lower. If, on the other hand, the critical temperature is higher, the total amount of SiO in the envelope becomes higher. In that case, the direct core growth stops at lower planet masses. 
\subsection{Adiabatic index and EOS}
The adiabatic index of \ce{SiO} based on the data from the NIST-JANAF Themochemical Tables is in the range of $\gamma = 1.25-1.4$ \citep{Janaf}. Therefore, a realistic adiabatic index of a gas mixture of \ce{H2}, \ce{He}, and \ce{SiO} is likely $\gamma<1.4$, which is the value used in this paper. We compared the envelope profile calculated using a realistic adiabatic index to the case of $\gamma=1.4$ and found no significant difference. This is because the latent heat term dominates the moist adiabatic gradient in \cref{eq:moistad}. \citet{2021Ormel} compared envelopes with nonideal, tabulated EOS values to the case of ideal EOS. They found no significant difference between the two models while the planet is still accreting pebbles. However, they found that the nonideal EOS plays an important role for the post disk evolution of the planet. 

\subsection{Energy transport}
We employed the classic approach to assume that, in areas where the envelope is stable against convection, energy is instead transported via radiation. However, the temperatures and pressures in the inner envelope might be high enough for conduction to become an effective way of energy transport \citep{2020VazanHelled,2022MisenerSchlichting,2023Misener}. We followed \citet{2023Misener} and compared the conductivity $\lambda_\mathrm{cond}$ to the equivalent term of radiative energy transport  
\begin{equation}
    \lambda_\mathrm{rad}=\frac{L}{4\pi r^2} \left(\pardev{T}{r}\right)_\mathrm{rad}=\frac{16 \sigma T^3 }{3\kappa \rho}.
\end{equation} 
Conduction becomes important when $\lambda_\mathrm{cond}/\lambda_\mathrm{rad}>1$, where $\lambda_\mathrm{cond} \approx 4\,\si{\watt \per \m\per\K}$ \citep{1983Stevenson}. For our parameters, this only happens deep in the envelope, where $q>q_\mathrm{max}$ and energy is transported efficiently by convection anyway. 

In addition, the diffusion of heat and vapor have different efficiency, which can lead to double-diffusive convection \citep{1960Stern,2011Rosenblum}. However, \citet{2017Leconte} found that condensation in a saturated medium also stabilizes the medium against double-diffusive convection. Therefore, we can assume that in regions that are stable against convection, energy is transported via radiation.  
\section{Summary and conclusion}
\label{sec:summary}
In this paper, we present a vapor equilibrium model for the envelope of an accreting rocky planet close to the water ice line. The model takes the enrichment of the envelope due to the evaporation of SiO from the magma ocean. The key assumption of this model is that the envelope is always in equilibrium with the underlying magma ocean. We then calculated the envelope structure of planets in the mass range $\Mp=0.1\,\Mearth$ to $\Mp = 3\,\Mearth$. Compared to pure \ce{H2}/He envelopes, the temperatures in the inner envelope of planets with $\Mp > 0.29\,\Mearth$ are higher if magma ocean evaporation is taken into account. This is due to the buildup of a mean molecular weight gradient, which stabilizes the transition from \ce{H2}-dominated to SiO-dominated gas against convection. Therefore, an inner radiative region with a steep temperature gradient forms. The region interior of the radiative zone of the planet is dominated in mass by gaseous SiO for $\Mp > 0.4\,\Mearth$. 

Almost all the SiO vapor is concentrated inside the inner radiative region and therefore protected from recycling flows that penetrate from the protoplanetary disk. As the planet grows, the saturated vapor pressure increases and more SiO vapor evaporates from the magma ocean. For planets with $\Mp \geq 2.0\,\Mearth$, the amount of SiO needed to keep the envelope saturated is larger than the incoming pebble flux. At this point, the accreted pebbles sublimate their SiO directly in the envelope and the luminosity profile of the envelope changes to reach vapor equilibrium at a lowered accretion luminosity.

Overall, we showed that our results are relatively independent of the pebble accretion rate and the detailed treatment of opacity in the envelope. However, envelopes where only gas opacity is considered tend to be slightly colder than envelopes where both dust and gas are considered as opacity sources. Similarly, for the same planet mass, a lower pebble accretion rate leads to lower temperatures in the envelope.   

The sublimation of pebbles in the envelope is an important aspect of the formation of planets by pebble accretion. In combination with other envelope processes such as recycling, the sublimation of volatile to moderately volatile elements can shape the composition of the planet \citep{2021Johansen,2023Steinmeyer,2023Wang}. At the same time, the enrichment of the envelope in \ce{H2O} and \ce{SiO} can also lead to increased gas accretion rates, which facilitates the formation of gas giants \citep{2015Venturini,2020Valletta,2021Ormel}. One of the most important consequences of pebble sublimation, however, is the buildup of a composition gradient in the envelope even at low planet masses. Future work is needed to study the long-term evolution of these envelopes for a wide range of parameters. In this work we focused on the accretion of silicates for planets located at $1\,\si{\au}$. A logical next step different is therefore to study the envelope structure of planets at larger orbital distances. At these distance, the accretion of water ice will play an important role. Therefore water vapor should be added to the model in the future.

Additionally, more work is needed to study the dynamics of these envelopes in 2-D or 3-D simulations. It is also important that the effects of SiO saturation are included in population synthesis models to better connect our understanding of planet formation to the observed exoplanet population. 
\begin{acknowledgements}
    We thank the anonymous reviewer for their useful comments. We thank Jérémy Leconte for helpful discussions on the topic of convective processes. A.J. acknowledges funding from the European Research Foundation (ERC Consolidator Grant 724687-PLANETESYS), the Knut and Alice Wallenberg Foundation (Wallenberg Scholar Grant 2019.0442), the Swedish Research Council (Project Grant 2018-04867), the Danish National Research Foundation (DNRF Chair Grant DNRF159), the Göran Gustafsson Foundation and the Carlsberg Foundation (Semper Ardens: Advance grant FIRSTATMO). This paper makes use of the following Python3 packages: numpy \citep{numpy} and matplotlib \citep{matplotlib}.
\end{acknowledgements}
\bibliographystyle{aa} 

\begin{thebibliography}{69}
\expandafter\ifx\csname natexlab\endcsname\relax\def\natexlab#1{#1}\fi
 \newcommand{\noop}[1]{}
\expandafter\ifx\csname natexlab\endcsname\relax\def\natexlab#1{#1}\fi

\bibitem[{{Ali-Dib} \& {Thompson}(2020)}]{2020Alidibthompson}
{Ali-Dib}, M. \& {Thompson}, C. 2020, \apj, 900, 96

\bibitem[{Alibert(2017)}]{2017Alibert}
Alibert, Y. 2017, Astronomy \& Astrophysics, 606, A69

\bibitem[{{Batygin} \& {Morbidelli}(2023)}]{2023BatyginMorbidelli}
{Batygin}, K. \& {Morbidelli}, A. 2023, Nature Astronomy, 7, 330

\bibitem[{{Bell} \& {Lin}(1994)}]{1994Bell}
{Bell}, K.~R. \& {Lin}, D.~N.~C. 1994, \apj, 427, 987

\bibitem[{{Bodenheimer} {et~al.}(2018){Bodenheimer}, {Stevenson}, {Lissauer},
  \& {D'Angelo}}]{2018Bodenheimer}
{Bodenheimer}, P., {Stevenson}, D.~J., {Lissauer}, J.~J., \& {D'Angelo}, G.
  2018, \apj, 868, 138

\bibitem[{Brouwers \& Ormel(2020)}]{brouwers2020planets}
Brouwers, M. \& Ormel, C. 2020, Astronomy \& Astrophysics, 634, A15

\bibitem[{{Brouwers} {et~al.}(2018){Brouwers}, {Vazan}, \&
  {Ormel}}]{2018Brouwers}
{Brouwers}, M.~G., {Vazan}, A., \& {Ormel}, C.~W. 2018, \aap, 611, A65

\bibitem[{{Bryson} {et~al.}(2021){Bryson}, {Kunimoto}, {Kopparapu}, {Coughlin},
  {Borucki}, {Koch}, {Aguirre}, {Allen}, {Barentsen}, {Batalha}, {Berger},
  {Boss}, {Buchhave}, {Burke}, {Caldwell}, {Campbell}, {Catanzarite},
  {Chandrasekaran}, {Chaplin}, {Christiansen}, {Christensen-Dalsgaard},
  {Ciardi}, {Clarke}, {Cochran}, {Dotson}, {Doyle}, {Duarte}, {Dunham},
  {Dupree}, {Endl}, {Fanson}, {Ford}, {Fujieh}, {Gautier}, {Geary},
  {Gilliland}, {Girouard}, {Gould}, {Haas}, {Henze}, {Holman}, {Howard},
  {Howell}, {Huber}, {Hunter}, {Jenkins}, {Kjeldsen}, {Kolodziejczak},
  {Larson}, {Latham}, {Li}, {Mathur}, {Meibom}, {Middour}, {Morris}, {Morton},
  {Mullally}, {Mullally}, {Pletcher}, {Prsa}, {Quinn}, {Quintana}, {Ragozzine},
  {Ramirez}, {Sanderfer}, {Sasselov}, {Seader}, {Shabram}, {Shporer}, {Smith},
  {Steffen}, {Still}, {Torres}, {Troeltzsch}, {Twicken}, {Uddin}, {Van Cleve},
  {Voss}, {Weiss}, {Welsh}, {Wohler}, \& {Zamudio}}]{2021Bryson}
{Bryson}, S., {Kunimoto}, M., {Kopparapu}, R.~K., {et~al.} 2021, \aj, 161, 36

\bibitem[{{Chase}(1998)}]{Janaf}
{Chase}, M.~W. 1998, {NIST}-{JANAF} thermochemical tables (Fourth edition.
  Washington, DC : American Chemical Society ; New York : American Institute of
  Physics for the National Institute of Standards and Technology, 1998.)

\bibitem[{{Connolly}(2016)}]{2016Connolly}
{Connolly}, J. A.~D. 2016, Journal of Geophysical Research (Planets), 121, 1641

\bibitem[{{Debras} \& {Chabrier}(2019)}]{2019DebrasChabrierJuno}
{Debras}, F. \& {Chabrier}, G. 2019, \apj, 872, 100

\bibitem[{{Dressing} \& {Charbonneau}(2013)}]{2013Dressing}
{Dressing}, C.~D. \& {Charbonneau}, D. 2013, \apj, 767, 95

\bibitem[{{Fegley} \& {Schaefer}(2012)}]{2012FegleySchaefer}
{Fegley}, Bruce, J. \& {Schaefer}, L. 2012, arXiv e-prints, arXiv:1210.0270

\bibitem[{{Freedman} {et~al.}(2014){Freedman}, {Lustig-Yaeger}, {Fortney},
  {Lupu}, {Marley}, \& {Lodders}}]{2014Freedman}
{Freedman}, R.~S., {Lustig-Yaeger}, J., {Fortney}, J.~J., {et~al.} 2014, \apjs,
  214, 25

\bibitem[{{Gail}(1998)}]{Gail1998}
{Gail}, H.~P. 1998, \aap, 332, 1099

\bibitem[{{Guillot}(1995)}]{1995Guillot}
{Guillot}, T. 1995, Science, 269, 1697

\bibitem[{{Gurrutxaga} {et~al.}(2024){Gurrutxaga}, {Johansen}, {Lambrechts}, \&
  {Appelgren}}]{2024Gurrutxaga}
{Gurrutxaga}, N., {Johansen}, A., {Lambrechts}, M., \& {Appelgren}, J. 2024,
  \aap, 682, A43

\bibitem[{Harris {et~al.}(2020)Harris, Millman, van~der Walt, Gommers,
  Virtanen, Cournapeau, Wieser, Taylor, Berg, Smith, Kern, Picus, Hoyer, van
  Kerkwijk, Brett, Haldane, del R{\'{i}}o, Wiebe, Peterson,
  G{\'{e}}rard-Marchant, Sheppard, Reddy, Weckesser, Abbasi, Gohlke, \&
  Oliphant}]{numpy}
Harris, C.~R., Millman, K.~J., van~der Walt, S.~J., {et~al.} 2020, Nature, 585,
  357

\bibitem[{{Hayashi}(1981)}]{1981Hayashi}
{Hayashi}, C. 1981, Progress of Theoretical Physics Supplement, 70, 35

\bibitem[{{Helled} {et~al.}(2011){Helled}, {Anderson}, {Podolak}, \&
  {Schubert}}]{2011HelledUranusNeptune}
{Helled}, R., {Anderson}, J.~D., {Podolak}, M., \& {Schubert}, G. 2011, \apj,
  726, 15

\bibitem[{{Herbort} {et~al.}(2020){Herbort}, {Woitke}, {Helling}, \&
  {Zerkle}}]{2020Herbort}
{Herbort}, O., {Woitke}, P., {Helling}, C., \& {Zerkle}, A. 2020, \aap, 636,
  A71

\bibitem[{{Hughes}(2006)}]{2006Hughes}
{Hughes}, D. 2006, Journal of the British Astronomical Association, 116, 21

\bibitem[{Hunter(2007)}]{matplotlib}
Hunter, J.~D. 2007, Computing in Science \& Engineering, 9, 90

\bibitem[{{Ikoma} \& {Hori}(2012)}]{2012Ikoma}
{Ikoma}, M. \& {Hori}, Y. 2012, \apj, 753, 66

\bibitem[{Iosilevskiy {et~al.}(2014)Iosilevskiy, Gryaznov, \&
  Solovev}]{2014Iosilevskiy}
Iosilevskiy, I., Gryaznov, V., \& Solovev, A. 2014, HIGH TEMPERATURES-HIGH
  PRESSURES, 43, 227, 10th International Workshop on Subsecond Thermophysics,
  Karlsruhe, GERMANY, JUN 26-28, 2013

\bibitem[{{Johansen} \& {Lacerda}(2010)}]{2010Johansen}
{Johansen}, A. \& {Lacerda}, P. 2010, \mnras, 404, 475

\bibitem[{{Johansen} {et~al.}(2015){Johansen}, {Mac Low}, {Lacerda}, \&
  {Bizzarro}}]{2015Johansen}
{Johansen}, A., {Mac Low}, M.-M., {Lacerda}, P., \& {Bizzarro}, M. 2015,
  Science Advances, 1, 1500109

\bibitem[{{Johansen} \& {Nordlund}(2020)}]{2020JohansenNordlund}
{Johansen}, A. \& {Nordlund}, {\r{A}}. 2020, \apj, 903, 102

\bibitem[{{Johansen} {et~al.}(2021){Johansen}, {Ronnet}, {Bizzarro},
  {Schiller}, {Lambrechts}, {Nordlund}, \& {Lammer}}]{2021Johansen}
{Johansen}, A., {Ronnet}, T., {Bizzarro}, M., {et~al.} 2021, Science Advances,
  7 [\eprint[arXiv]{2102.08611}]

\bibitem[{{Johansen} {et~al.}(2023){Johansen}, {Ronnet}, {Schiller}, {Deng}, \&
  {Bizzarro}}]{2023Johansen}
{Johansen}, A., {Ronnet}, T., {Schiller}, M., {Deng}, Z., \& {Bizzarro}, M.
  2023, \aap, 671, A75

\bibitem[{{Kippenhahn} {et~al.}(2013){Kippenhahn}, {Weigert}, \&
  {Weiss}}]{2013Kippenhahn}
{Kippenhahn}, R., {Weigert}, A., \& {Weiss}, A. 2013, {Stellar Structure and
  Evolution} (Springer Berlin, Heidelberg)

\bibitem[{{Kraus} {et~al.}(2012){Kraus}, {Stewart}, {Swift}, {Bolme}, {Smith},
  {Hamel}, {Hammel}, {Spaulding}, {Hicks}, {Eggert}, \& {Collins}}]{2012Kraus}
{Kraus}, R.~G., {Stewart}, S.~T., {Swift}, D.~C., {et~al.} 2012, Journal of
  Geophysical Research (Planets), 117, E09009

\bibitem[{{Kurokawa} \& {Tanigawa}(2018)}]{2018KurokawaTanigawa}
{Kurokawa}, H. \& {Tanigawa}, T. 2018, \mnras, 479, 635

\bibitem[{{Lambrechts} \& {Johansen}(2012)}]{Lambrechts2012}
{Lambrechts}, M. \& {Johansen}, A. 2012, Astronomy and Astrophysics, 544, A32

\bibitem[{{Lambrechts} {et~al.}(2014){Lambrechts}, {Johansen}, \&
  {Morbidelli}}]{2014Lambrechts}
{Lambrechts}, M., {Johansen}, A., \& {Morbidelli}, A. 2014, \aap, 572, A35

\bibitem[{{Lambrechts} \& {Lega}(2017)}]{2017LambrechtsLega}
{Lambrechts}, M. \& {Lega}, E. 2017, \aap, 606, A146

\bibitem[{{Lambrechts} {et~al.}(2019){Lambrechts}, {Morbidelli}, {Jacobson},
  {Johansen}, {Bitsch}, {Izidoro}, \& {Raymond}}]{2019Lambrechts}
{Lambrechts}, M., {Morbidelli}, A., {Jacobson}, S.~A., {et~al.} 2019, \aap,
  627, A83

\bibitem[{{Leconte} {et~al.}(2017){Leconte}, {Selsis}, {Hersant}, \&
  {Guillot}}]{2017Leconte}
{Leconte}, J., {Selsis}, F., {Hersant}, F., \& {Guillot}, T. 2017, \aap, 598,
  A98

\bibitem[{{Ledoux}(1947)}]{1947Ledoux}
{Ledoux}, P. 1947, \apj, 105, 305

\bibitem[{{Lee} {et~al.}(2014){Lee}, {Chiang}, \& {Ormel}}]{2014Lee}
{Lee}, E.~J., {Chiang}, E., \& {Ormel}, C.~W. 2014, \apj, 797, 95

\bibitem[{{Levison} {et~al.}(2015){Levison}, {Kretke}, {Walsh}, \&
  {Bottke}}]{2015Levison}
{Levison}, H.~F., {Kretke}, K.~A., {Walsh}, K.~J., \& {Bottke}, W.~F. 2015,
  Proceedings of the National Academy of Science, 112, 14180

\bibitem[{{Lodders}(2003)}]{2003Lodders}
{Lodders}, K. 2003, \apj, 591, 1220

\bibitem[{{Markham} {et~al.}(2022){Markham}, {Guillot}, \&
  {Stevenson}}]{2022Markham}
{Markham}, S., {Guillot}, T., \& {Stevenson}, D. 2022, \aap, 665, A12

\bibitem[{{Marley} {et~al.}(1995){Marley}, {G{\'o}mez}, \&
  {Podolak}}]{1995Marley}
{Marley}, M.~S., {G{\'o}mez}, P., \& {Podolak}, M. 1995, \jgr, 100, 23349

\bibitem[{{Melosh}(2007)}]{2007Melosh}
{Melosh}, H.~J. 2007, \maps, 42, 2079

\bibitem[{{Militzer} \& {Hubbard}(2024)}]{2024MilitzerHubbard}
{Militzer}, B. \& {Hubbard}, W.~B. 2024, arXiv e-prints, arXiv:2401.12166

\bibitem[{{Misener} \& {Schlichting}(2022)}]{2022MisenerSchlichting}
{Misener}, W. \& {Schlichting}, H.~E. 2022, \mnras, 514, 6025

\bibitem[{{Misener} {et~al.}(2023){Misener}, {Schlichting}, \&
  {Young}}]{2023Misener}
{Misener}, W., {Schlichting}, H.~E., \& {Young}, E.~D. 2023, \mnras, 524, 981

\bibitem[{{Morbidelli} \& {Nesvorny}(2012)}]{2012MorbidelliNesvorny}
{Morbidelli}, A. \& {Nesvorny}, D. 2012, \aap, 546, A18

\bibitem[{{Olson} {et~al.}(2022){Olson}, {Sharp}, \& {Garai}}]{2022Olson}
{Olson}, P., {Sharp}, Z., \& {Garai}, S. 2022, Earth and Planetary Science
  Letters, 587, 117537

\bibitem[{{Ormel} \& {Klahr}(2010)}]{OrmelKlahr2010}
{Ormel}, C.~W. \& {Klahr}, H.~H. 2010, \aap, 520, A43

\bibitem[{{Ormel} {et~al.}(2021){Ormel}, {Vazan}, \& {Brouwers}}]{2021Ormel}
{Ormel}, C.~W., {Vazan}, A., \& {Brouwers}, M.~G. 2021, \aap, 647, A175

\bibitem[{{Podolak} {et~al.}(2000){Podolak}, {Podolak}, \&
  {Marley}}]{2000Podolak}
{Podolak}, M., {Podolak}, J.~I., \& {Marley}, M.~S. 2000, \planss, 48, 143

\bibitem[{{Pollack} {et~al.}(1994){Pollack}, {Hollenbach}, {Beckwith},
  {Simonelli}, {Roush}, \& {Fong}}]{1994Pollack}
{Pollack}, J.~B., {Hollenbach}, D., {Beckwith}, S., {et~al.} 1994, \apj, 421,
  615

\bibitem[{{Raymond} {et~al.}(2020){Raymond}, {Izidoro}, \&
  {Morbidelli}}]{2020Raymond}
{Raymond}, S.~N., {Izidoro}, A., \& {Morbidelli}, A. 2020, in Planetary
  Astrobiology, ed. V.~S. {Meadows}, G.~N. {Arney}, B.~E. {Schmidt}, \& D.~J.
  {Des Marais}, 287

\bibitem[{{Raymond} {et~al.}(2009){Raymond}, {O'Brien}, {Morbidelli}, \&
  {Kaib}}]{2009Raymond}
{Raymond}, S.~N., {O'Brien}, D.~P., {Morbidelli}, A., \& {Kaib}, N.~A. 2009,
  \icarus, 203, 644

\bibitem[{{Rogers}(2015)}]{2015Rogers}
{Rogers}, L.~A. 2015, \apj, 801, 41

\bibitem[{{Ros} {et~al.}(2019){Ros}, {Johansen}, {Riipinen}, \&
  {Schlesinger}}]{2019Ros}
{Ros}, K., {Johansen}, A., {Riipinen}, I., \& {Schlesinger}, D. 2019, \aap,
  629, A65

\bibitem[{{Rosenblum} {et~al.}(2011){Rosenblum}, {Garaud}, {Traxler}, \&
  {Stellmach}}]{2011Rosenblum}
{Rosenblum}, E., {Garaud}, P., {Traxler}, A., \& {Stellmach}, S. 2011, \apj,
  731, 66

\bibitem[{{Schaefer} \& {Fegley}(2009)}]{2009Schaefer}
{Schaefer}, L. \& {Fegley}, B. 2009, \apjl, 703, L113

\bibitem[{{Schaefer} {et~al.}(2012){Schaefer}, {Lodders}, \&
  {Fegley}}]{2012Schaefer}
{Schaefer}, L., {Lodders}, K., \& {Fegley}, B. 2012, \apj, 755, 41

\bibitem[{{Silburt} {et~al.}(2015){Silburt}, {Gaidos}, \& {Wu}}]{2015Silburt}
{Silburt}, A., {Gaidos}, E., \& {Wu}, Y. 2015, \apj, 799, 180

\bibitem[{{Steinmeyer} {et~al.}(2023){Steinmeyer}, {Woitke}, \&
  {Johansen}}]{2023Steinmeyer}
{Steinmeyer}, M.-L., {Woitke}, P., \& {Johansen}, A. 2023, \aap, 677, A181

\bibitem[{{Stern}(1960)}]{1960Stern}
{Stern}, M.~E. 1960, Tellus, 12, 172

\bibitem[{Stevenson {et~al.}(1983)Stevenson, Spohn, \&
  Schubert}]{1983Stevenson}
Stevenson, D.~J., Spohn, T., \& Schubert, G. 1983, Icarus, 54, 466

\bibitem[{{Tachibana} {et~al.}(2002){Tachibana}, {Tsuchiyama}, \&
  {Nagahara}}]{2002Tachibana}
{Tachibana}, S., {Tsuchiyama}, A., \& {Nagahara}, H. 2002, \gca, 66, 713

\bibitem[{{Tsuchiyama} {et~al.}(1999){Tsuchiyama}, {Tachibana}, \&
  {Takahashi}}]{1999Tsuchiyama}
{Tsuchiyama}, A., {Tachibana}, S., \& {Takahashi}, T. 1999, \gca, 63, 2451

\bibitem[{{Valletta} \& {Helled}(2020)}]{2020Valletta}
{Valletta}, C. \& {Helled}, R. 2020, \apj, 900, 133

\bibitem[{{Vazan} \& {Helled}(2020)}]{2020VazanHelled}
{Vazan}, A. \& {Helled}, R. 2020, \aap, 633, A50

\bibitem[{{Vazan} \& {Ormel}(2023)}]{2023VazanOrmel}
{Vazan}, A. \& {Ormel}, C.~W. 2023, \aap, 676, L8

\bibitem[{{Venturini} {et~al.}(2015){Venturini}, {Alibert}, {Benz}, \&
  {Ikoma}}]{2015Venturini}
{Venturini}, J., {Alibert}, Y., {Benz}, W., \& {Ikoma}, M. 2015, \aap, 576,
  A114

\bibitem[{{Venturini} \& {Helled}(2020)}]{2020VenturiniHelled}
{Venturini}, J. \& {Helled}, R. 2020, \aap, 634, A31

\bibitem[{{Visscher} \& {Fegley}(2013)}]{2013VisscherFegley}
{Visscher}, C. \& {Fegley}, Bruce, J. 2013, \apjl, 767, L12

\bibitem[{{Wahl} {et~al.}(2017){Wahl}, {Hubbard}, {Militzer}, {Guillot},
  {Miguel}, {Movshovitz}, {Kaspi}, {Helled}, {Reese}, {Galanti}, {Levin},
  {Connerney}, \& {Bolton}}]{2017WahlJuno}
{Wahl}, S.~M., {Hubbard}, W.~B., {Militzer}, B., {et~al.} 2017, \grl, 44, 4649

\bibitem[{{Wang} {et~al.}(2023){Wang}, {Ormel}, {Huang}, \&
  {Kuiper}}]{2023Wang}
{Wang}, Y., {Ormel}, C.~W., {Huang}, P., \& {Kuiper}, R. 2023, \mnras, 523,
  6186

\bibitem[{{Wolfgang} \& {Lopez}(2015)}]{2015Wolfgang}
{Wolfgang}, A. \& {Lopez}, E. 2015, \apj, 806, 183

\bibitem[{{Xiao} \& {Stixrude}(2018)}]{2018XiaoStixrude}
{Xiao}, B. \& {Stixrude}, L. 2018, Proceedings of the National Academy of
  Science, 115, 5371

\bibitem[{{Zeng} {et~al.}(2019){Zeng}, {Jacobsen}, {Sasselov}, {Petaev},
  {Vanderburg}, {Lopez-Morales}, {Perez-Mercader}, {Mattsson}, {Li}, {Heising},
  {Bonomo}, {Damasso}, {Berger}, {Cao}, {Levi}, \& {Wordsworth}}]{2019Zeng}
{Zeng}, L., {Jacobsen}, S.~B., {Sasselov}, D.~D., {et~al.} 2019, Proceedings of
  the National Academy of Science, 116, 9723

\end{thebibliography}

\end{document}